\documentclass[final,3p,times,twocolumn]{elsarticle}
\usepackage{float}
\usepackage{graphics}

\journal{Communications in Nonlinear Science and Numerical Simulation}

\usepackage{graphicx}
 \usepackage{epsfig}

\begin{document}
\def\ds{\displaystyle}
\begin{frontmatter}

\title{ Nonlinear waves in an anti-Hermitian lattice with cubic nonlinearity}
\author{S. Tombuloglu $^{\dagger}$, C. Yuce}
\address{ $^{\dagger}$ Saglik Hizmetleri MYO, Kirklareli University, Kirklareli, Turkey\\
Department of Physics, Faculty of Science, Eskisehir Technical University, Eskisehir, Turkey}
\ead{cyuce@eskisehir.edu.tr} \fntext[label2]{}
\begin{abstract}
In an anti-Hermitian linear system, all energy eigenvalues are purely imaginary and the corresponding eigenvectors are orthogonal. This implies that no stationary state is available in such systems. We consider an anti-Hermitian lattice with cubic nonlinearity and explore novel nonlinear stationary modes. We discuss that relative population is conserved in a nonreciprocal tight binding lattice with periodical boundary conditions as opposed to parity-time (PT) symmetric lattices. We study nonlinear nonrecipocal dimer, triple and quadrimer models and construct stationary nonlinear modes.
\end{abstract}


\end{frontmatter}


\section{Introduction}

Two decades ago, Bender and Boettcher showed that a non-Hermitian quantum mechanical Hamiltonian with simultaneous $\mathcal{P}$ and $\mathcal{T}$ symmetries can have real energy eigenvalues \cite{bender}. Here, the $\mathcal{P}$ stands for the parity defined as the reflection against the origin of the coordinate system and the $\mathcal{T}$ stands for the the time reversal. In a decade, extension of $\mathcal{PT}$ symmetric systems to include nonlinearity was suggested \cite{mslmn}. It was shown that steady states can exist as continuous families in a nonlinear $\mathcal{PT}$-symmetric  system. Over the last decade, nonlinear dynamics of non-Hermitian systems has attracted a great deal of attention \cite{rev01,rev02}. Nonlinear dissipative systems are not only of theoretical interest. The theoretical predictions can be tested in experiments, which can be realized with current technology using the similarity between the Schrodinger equation and the paraxial wave equation \cite{deney01}. Of special importance is $\mathcal{PT}$ symmetric discrete lattices governed by discrete nonlinear Schrodinger (dNLS) type equations. In these systems, the $\mathcal{PT}$-symmetry is obtained by judiciously inserting balanced gain and loss to lattice sites, which are coupled to their neighbouring sites. It was shown that the $\mathcal{PT}$-symmetry gets fragile as the number of lattice sites is increased \cite{fragile}. The dimer model, which is the simplest $\mathcal{PT}$-symmetric nonlinear discrete lattice was shown to be integrable \cite{dimer1,dimer2,dimer3,dimer4,dimer5,dimer6,dimer7}. More complex systems such as the nonlinear $\mathcal{PT}$-symmetric trimer \cite{trimer1,trimer2}, quadrimer \cite{quad003}, 2D $\mathcal{PT}$-symmetric plaquettes \cite{plaqu}, generalized nonlinear $\mathcal{PT}$-symmetric lattice \cite{nongener,nongener2,nonpt1}, non-Hermitian nonlinear necklaces \cite{necklaces}, asymmetric wave propagation \cite{awp1} and stability of discrete solitons \cite{insta} were alo studied. The dynamics of such systems were explored \cite{dyn00,dyn01}. In a recent study, a nonlinear system with charge-parity-symmetric dimers was considered \cite{CP}. A $\mathcal{PT}$-symmetric coupler, with additional gain and loss proportional to nonlinear terms has been studied in \cite{nlgain}.\\
The study of non-Hermitian nonlinear systems has been mainly restricted to systems with gain and loss impurities in the literature. There are some other ways to induce non-Hermiticity other than gain and loss impurities \cite{nlgain}. One of them is to introduce non-reciprocity arising from asymmetrical hopping amplitudes in a lattice \cite{exp1,cem1,cem2}. Although $\mathcal{PT}$-symmetric nonlinear lattices have been studied by many authors \cite{rev01,rev02}, the nonreciprocal nonlinear lattices are mostly unexplored. In this paper, we start with a linear tight-binding lattice with asymmetrical hopping amplitudes. We specifically study anti-Hermitian system and then generalize it to include nonlinear interaction. In an anti-Hermitian linear system, i. e., $\ds{\mathcal{H}=-\mathcal{H}^{\dagger} }$, eigenvectors are orthogonal, which implies that one can construct Hilbert space. However, all energy eigenvalues are purely imaginary. Therefore no stable state is available. We show that nonlinearity leads to stable modes. We explore some novel effects in nonlinear dimer such as the relative population conservation. We study nonlinear nonreciprocal dimer, triple and quadrimer models and construct stationary nonlinear modes.

\section{Dimer}

Consider a nonreciprocal dimer, which has asymmetric hopping (tunnelling) amplitudes. In the linear case, the system Hamiltonian is anti-Hermitian, which implies that corresponding eigenvalues are purely imaginary. In the presence of conservative Kerr nonlinear interaction, the dynamical equations are of the form of discrete nonlinear Schrodinger equations
\begin{eqnarray}\label{o987jj2}
i~\dot{u}_1=~~\kappa ~u_2+g~ |u_1|^2~ u_1 \nonumber\\
i~\dot{u}_2=-\kappa ~u_1+g~ |u_2|^2 ~u_2
\end{eqnarray}
where the overdot denotes time derivation, $g$ is the nonlinearity strength and the hopping amplitudes in the forward and backward directions have opposite signs with magnitude $\kappa$. This system can describe light propagation through two waveguides coupled to each other asymmetrically. In this case, $u_1$ and $u_2$ become modal field amplitudes and time parameter $t$ is replaced by the propagation distance.\\
The fundamental difference between the nonreciprocal dimer and its Hermitian counterpart is that the former one is lacking the conservation of the total intensity. Fortunately, the nonreciprocal dimer has its own characteristic conserved quantity. It is the relative population (intensity), $\ds{|u_1|^2-|u_2|^2}$. Using the equations (\ref{o987jj2}), one can show that the relative population is constant in time in the nonreciprocal dimer, regardless of initial states and the nonlinear interaction strength
\begin{equation}\label{vahdeklas}
\frac{d(|u_1|^2-|u_2|^2)}{dt}=0
\end{equation}
This implies that no intensity oscillation occurs between the two sites in both linear and nonlinear cases. The two sites can have only time-dependent phase difference. As we will see below, the system can either grow or decay as a whole with no particle transfer between the sites in the linear case, $g=0$. As opposed to the linear system, the total intensity can also be a constant or oscillate in time while the relative intensity remains constant in the nonlinear case. Note that neither the total intensity nor the relative intensity is conserved in time in non-Hermitian $\mathcal{PT}$-symmetric nonlinear dimer. \\
Let us first study the linear system $g=0$. Suppose that the system has initially equal probability on both sides. Then the only difference can come from the phase difference. Therefore, we write
\begin{equation}\label{cozumas}
u_2=e^{i\Lambda(t)} u_1
\end{equation}
where the time-dependent function $\Lambda(t)$ is the relative phase. If we substitute this expression into (\ref{o987jj2}), we get
\begin{equation}\label{cos7hnb}
\dot{\Lambda}  = ~2 \kappa\cos{\Lambda} 
\end{equation}
The stationary solution for which the phase difference between the two sites is constant can be obtained by setting $\dot{\Lambda}=0$. The corresponding solution reads $\ds{ e^{2i\Lambda} =-1    }$, which implies that $\ds{u_2(t)={\mp}i u_1(t)}$. In this case, the solution is given by
\begin{equation}\label{cbdfjsams}
(u_1,u_2)=N e^{{\mp}t} (1,{\mp}i) 
\end{equation}
where $N$ is a constant. One can see that the corresponding energy eigenvalues are purely imaginary. In other words, the two eigenstates either grow or decay in time depending on the sign of energy eigenvalues. We emphasize that these two eigenstates are orthogonal to each other and hence the Hilbert space exists in the system as opposed to the most non-Hermitian systems. Therefore, any initial state can be expanded in terms of these two eigenstates, which implies that no constant intensity solution is available in the linear case since any initial state either grows or decays in time.\\
Let us now study non-stationary solutions of (\ref{cos7hnb}). Assume that the initial condition is given by $\ds{\Lambda(t=0)=\Lambda_0}$, where $-\pi\leq\Lambda_0\leq\pi$. It is interesting to see numerically that the solution rapidly comes to a fixed number such that $\ds{ \cos{\Lambda}=0   }$ (and $\ds{ \sin{\Lambda}=1   }$ ). This means that the final state becomes the eigenstate with $+$ sign in (\ref{cbdfjsams}), regardless of initial states. This is because of the fact that the eigenstate with $-$ sign decays in time and the other one becomes the remaining eigenstate after a while. To check our discussion, we numerically solve (\ref{o987jj2}) at $g=0$ for the initial values $\ds{u_1(0)=1}$ and $\ds{u_2(0)=i\lambda}$, where $\ds{\lambda}$ is a real valued constant. Suppose that $\ds{\lambda}$ is very close to $-1$. We numerically see that the solution decays first and then grows in time unboundedly. The state at large times is the one with $+$ sign.\\
As opposed to the linear system where only growing/decaying solutions are available, the nonlinearity provides constant intensity solutions. To study them, we assume $\ds{u_1(t)=e^{-iE t} U_1}$ and $\ds{u_2(t)=e^{-iE t} U_2}$, where $\ds{E}$ is a real continuous parameter, $U_1$ and $U_2$ are all time-independent constants. Therefore, we get
\begin{eqnarray}\label{580sb2}
E~U_1=~~\kappa~ U_2+g~ |U_1|^2 ~U_1 \nonumber\\
E~U_2=-\kappa ~U_1+g~ |U_2|^2~ U_2
\end{eqnarray}
There are two coupled third-degree polynomials so there exists $9$ solutions. One of them is a trivial solution $(U_1=0,U_2=0)$. Therefore, we look for $8$ solutions. Note that they satisfy the symmetry condition (if a particular $(U_1,U_2)$ satisfies the equation (\ref{580sb2}), then $(-U_1,-U_2)$ is also a solution). Note also that no stationary solution with equally distributed intensity between the two sites ($\ds{U_1=U_2}$) exists as opposed to both the Hermitian and $\mathcal{PT}$  symmetric nonlinear dimer. \\
Let us look for solutions where $U_1$ and $U_2$ are real numbers. This implies that phase current doesn't arise between the sites. One can see from (\ref{580sb2}) that $\ds{U_2=\kappa^{-1}(E-g~U_1^2) ~U_1}$. We can then analytically obtain them
\begin{eqnarray}\label{5765utrfnc2}
U_1^2=\frac{    3E-\sqrt{E^2-8\kappa^2} \mp  \sqrt{8\kappa^2+2E    ( E+\sqrt{E^2-8\kappa^2}      ) } }{ 4g }    \nonumber\\
U_1^2=\frac{    3E+\sqrt{E^2-8\kappa^2} \mp  \sqrt{8\kappa^2+2E    ( E-\sqrt{E^2-8\kappa^2}      ) } }{ 4g }   
\end{eqnarray}
These are nonlinear modes since they are absent in the linear model, $g=0$. We get continuous families of nonlinear modes by varying the free real-valued parameter $\ds{E}$. No such solutions are available if $\ds{|E|<\sqrt{8}  \kappa}$. In fact, they exist only for $\ds{E\geq\sqrt{8} \kappa}$ for $g>0$ and $\ds{E\leq\sqrt{8}  \kappa }$ for $g<0$. In Fig.1.a, we plot these modes as a function of $\ds{E}$ for $g=1$ and $\kappa=1$. Below we study only the modes with $U_1>0$ since the modes are symmetric with respect to $E$-axis. As can be seen from the figure, the modes are bifurcating from zero amplitude at $\ds{E=\sqrt{8}}$. There are $2$ (positive-valued) modes at the branch point $E=\sqrt{8}$ and $4$ such modes when $E>\sqrt{8}$. The amplitude of one of those modes decreases with increasing $E$ while the amplitudes of all other modes  tend to infinity as $\ds{E}$ increases. Let us now study the linear stability of these nonlinear modes by linearizing the nonlinear equations. Adding small amplitude modulation to a stationary solution changes the initial value of relative population, which in turn remains constant in time. The small amplitude modulation can either cause the system to either grow unboundedly or oscillate in time as a whole. This has no analogue in both Hermitian and $\mathcal{PT}$-symmetric nonlinear dimers. We say that the solution is unstable if it grows unboundedly as a whole. We perform numerical computation to study stability for the nonlinear modes with $U_1>0$. We numerically see that one of the $4$ modes (the gray curves in the Fig.1) is not stable against small amplitude modulation as it grows exponentially in time. Fortunately, no such an exponential growth occurs for all other modes under small amplitude modulation. Instead we see that they make small amplitude oscillations (breathing) such that $\ds{ |u_2|^2- |u_1|^2}$ is constant in time. The amplitude of the oscillation is highest when $U_1$ is the highest.   \\ 
We have assumed that $U_1$ and $U_2$ are all real numbers. If one of them is real while the other is purely imaginary, then $\ds{E}$ becomes complex valued. In this case, the nonlinear wave either grows without bound or the total intensity oscillates in time (breathing). Similar dynamics can also be seen when $U_1$ and $U_2$ are real-valued but not equal to their stationary values given in (\ref{5765utrfnc2}). The type of dynamics is determined sensitively by the initial intensities on both sites and the nonlinearity strength. As an example, let us consider the initial values $u_1(0)=u_2(0)=1$. The relative population remains zero at any time and hence it is enough to plot $|u_1(t)|$ since $|u_1(t)|=|u_2(t)|$. The Fig. 2 plots it for various values of $g$. In the linear case $g=0$, $u_1(t)$ grows exponentially in time as expected. The growth rate decreases with increasing $|g|$. At a critical value of $g$, the growth is prevented and the system has oscillating behaviour, where the amplitude of oscillations decreases with $|g|$. For large values of $|g|$, the system can be considered as almost stationary since the total intensities oscillate with very small amplitudes. To this end, we note that nonreciprocal evolution of initial state does not occur in our system (due to the conservation of relative population) as it does in the $\mathcal{PT}$-symmetric nonlinear dimer \cite{dimer1}.
\begin{figure}[t]
\includegraphics[width=6cm]{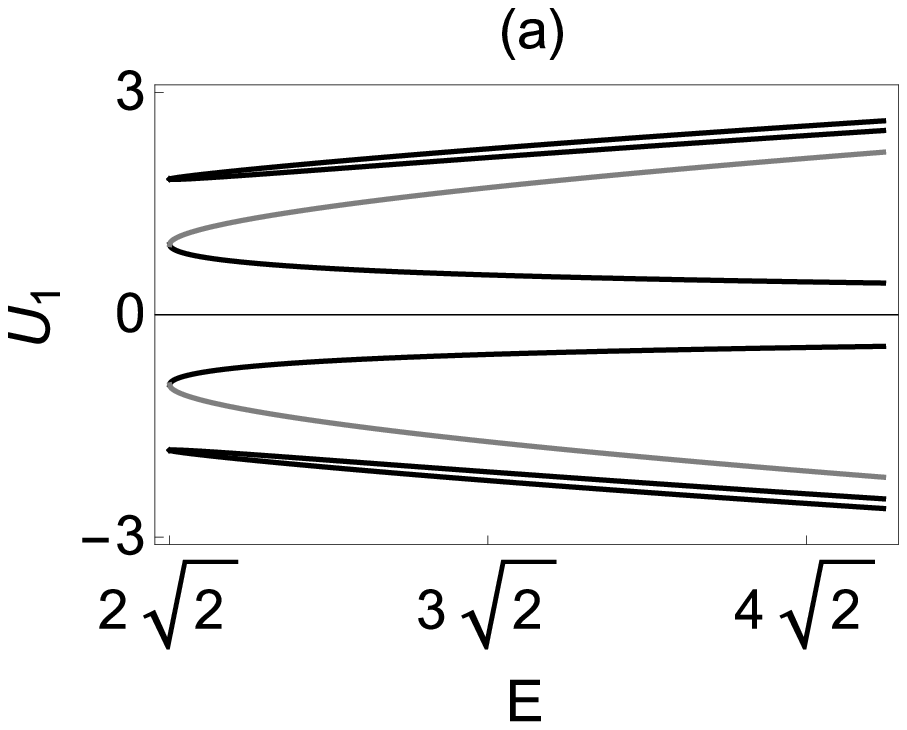}
\includegraphics[width=6cm]{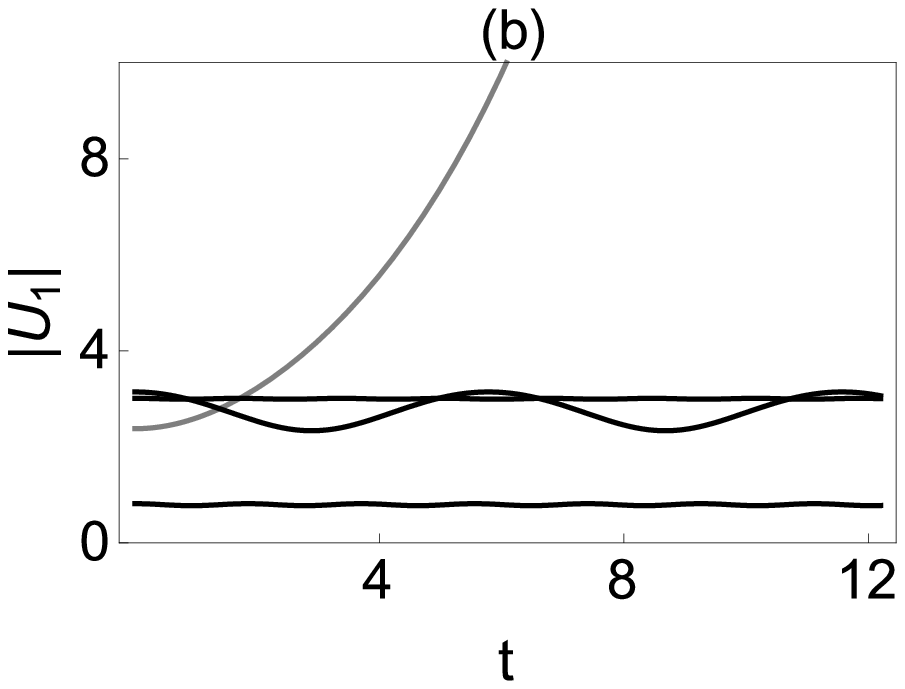}
\caption{ The field amplitude $U_1$ (\ref{5765utrfnc2}) versus energy $E$ (a) for a nonlinear nonreciprocal dimer. The parameters are given by $g=1$ and $\kappa=1$. No such solutions are available if $\ds{E<2\sqrt{2}}$. The curve in gray is unstable against small amplitude modulation and grows unboundedly in time. In (b), arbitrary small modulation is introduced to the solution at $E=4$ and the absolute value of the field amplitude $|U_1|$ is plotted as a function of time $t$. The one with the gray curve grows in time while the other ones oscillate in time.}
\end{figure}
\begin{figure}[t]
\includegraphics[width=6cm]{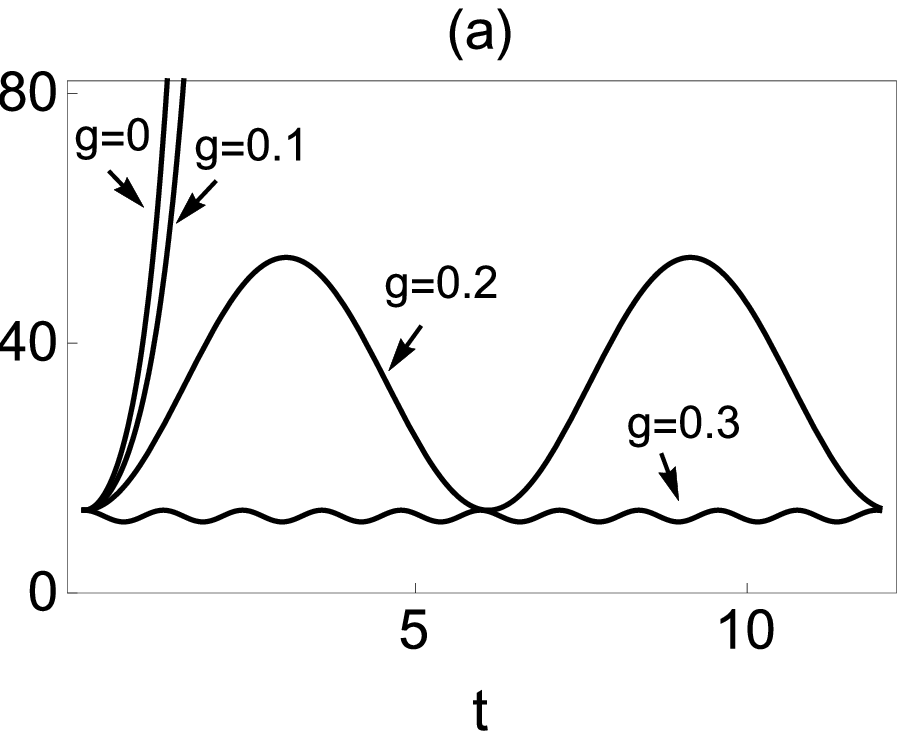}
\includegraphics[width=6cm]{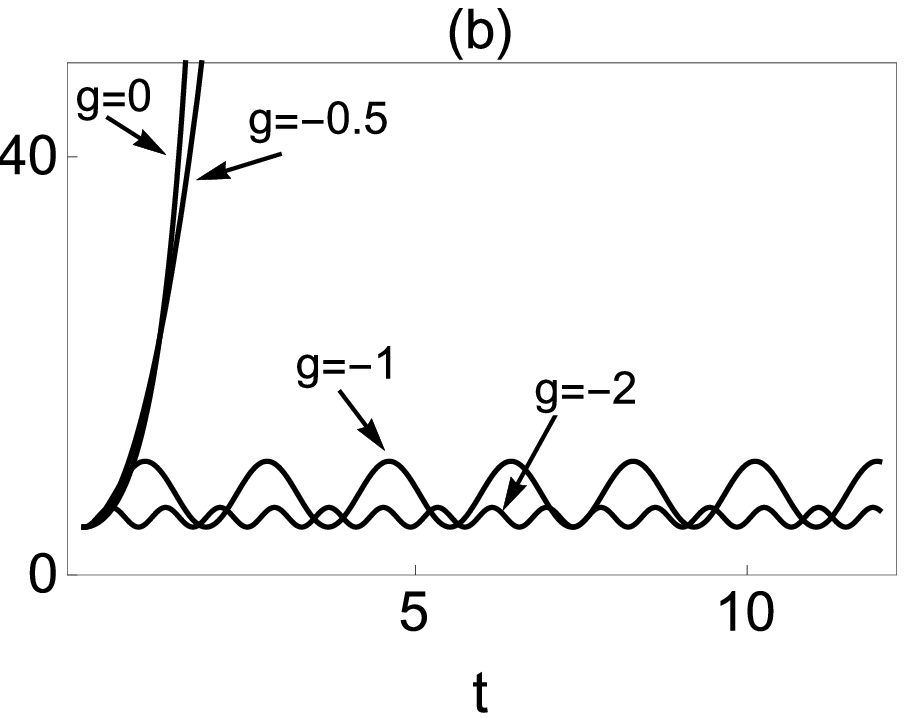}
\caption{ $|u_1|$ as a function of time for positive (a) and negative values of nonlinearity strength (b) at $\kappa=1$. We use the initial values $u_1(0)=u_2(0)=1$ for the numerical solution of (\ref{o987jj2}). The amplitude grows unboundedly in the linear system $g=0$. As $|g|$ is increased, oscillatory behaviours can be seen. The oscillation amplitude decreases with $|g|$. Note that the relative amplitude remains constant in time for all cases.  }
\end{figure}

\section{Trimer}

Let us now study nonlinear trimer model with asymmetric hopping amplitudes. The dynamical equations for the system is given by
\begin{eqnarray}\label{1ofhap2}
i~\dot{u}_1= u_2-u_3+g~ |u_1|^2~ u_1 \nonumber\\
i~\dot{u}_2= u_3-u_1+g~ |u_2|^2 ~u_2 \nonumber\\
i~\dot{u}_3=u_1-u_2+g~ |u_3|^2 ~u_3
\end{eqnarray}
where we set $\ds{\kappa=1}$ for the sake of simplicity. As opposed to the nonreciprocal dimer, the relative probability among the sites is not conserved in time. The system is anti-Hermitian in the linear case $g=0$ and hence all eigenvalues are purely imaginary. Therefore no stationary solution is available at $g=0$. Fortunately the system can support stationary solutions in the presence of nonlinear term. Let us look for stationary solutions of the form $\ds{u_{i}(t)=e^{-iE t} U_i}$, where $\ds{E}$ and $U_i$ are all time-independent constants. Therefore, we get
\begin{eqnarray}\label{2085sia2}
E~U_1&=& U_2-U_3+g~ |U_1|^2~ U_1 \nonumber\\
E~U_2&=& U_3-U_1+g~ |U_2|^2 ~U_2 \nonumber\\
E~U_3&=&U_1-U_2+g~ |U_3|^2 ~U_3 \
\end{eqnarray}
These equations admit at most $3^3=27$ solutions since there are $3$ coupled third-degree polynomials. We numerically solve them and plot specifically $U_1$ as a function of $E$ at $g=1$ in the Fig 3.a. Similar plots for $U_2$ and $U_3$ can also be obtained. As can be seen, stationary modes exist for all values of $E>0$ and the number of modes increases with the free parameter $E$. We see that the amplitudes $U_1$ at a given large values of $E$ are very close to each other for most of the modes. Let us now obtain analytical expressions for some of the modes. Firstly, there is a trivial solution such that $U_1=U_2=U_3=0$. The other modes have symmetry $\{-U_1,-U_2,-U_3\}$ and hence it is enough to find the $13$ (non-symmetric) solutions. One of them has the uniform density and is given by
\begin{equation}\label{dq1502l2}
U_1=U_2=U_3=\sqrt{ \frac{E}{g}}
\end{equation}
This solution exists for any values of $E$. In fact, it is the only nontrivial solution at low values of $E$. The nonlinearity strength $g$ plays an important role on the dynamics of this uniform solution under amplitude modulation. If $|g|$ is close to $0$, then the solution is unstable against small amplitude modulation and the total amplitude grows exponentially in time. For larger values of $|g|$, we see that the total amplitude shows non-periodical oscillatory behavior. In Fig 3.b, we plot $P(t)=\sqrt{u_1^2+u_2^2+u_3^2}$ as a function of time for the initial condition is $u_1(0)=u_2(0)=1,u_3(0)=0.99$. As can be seen, the amplitude of the oscillation decreases with increasing $g$.\\
There exists $6$ other solutions for which one of the mode amplitudes is small for large values of $E$. We can then approximately construct them. Suppose first that $U_2$ is small, i. e., $U_2\approx 0$. Therefore we neglect the nonlinear term $g|U_2|^2U_2$ in (\ref{2085sia2}). Consequently, we get
\begin{equation}\label{d9paXl2}
U_2\approx\frac{U_{3}-U_{1}}{E}
\end{equation}
If we substitute this into the other two equations in (\ref{2085sia2}) and assume that $E>>0$, then the nonlinear trimer is effectively reduced to the nonlinear dimer
\begin{eqnarray}\label{2wisa1a2}
E~U_1&\approx & -U_3+g~ |U_1|^2~ U_1 \nonumber\\
E~U_3&\approx&~~U_1+g~ |U_3|^2 ~U_3 \
\end{eqnarray}
This has the form of the equation (\ref{580sb2}) and the corresponding solutions can be obtained using (\ref{5765utrfnc2}). The stability features of these solutions has already been discussed in the previous section. We have derived them under the assumption of small values of $U_2$ If we repeat our formalism by assuming that $U_1$ is small and  $U_3$ is small, we get $6$ such solutions. We numerically check that this approximation works very well. There are $6$ other non-symmetric solutions left. One can get their exact analytical forms. But it is cumbersome to write them here. We numerically find that they are stable under arbitrary small amplitude modulation as their total intensities don't grow exponentially in time.
\begin{figure}[t]
\includegraphics[width=6cm]{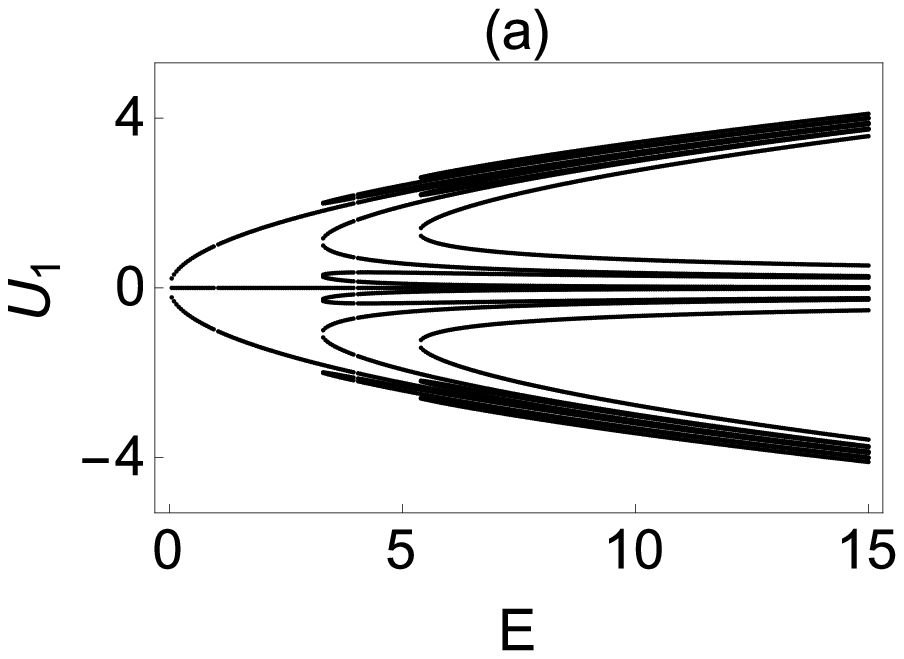}
\includegraphics[width=6cm]{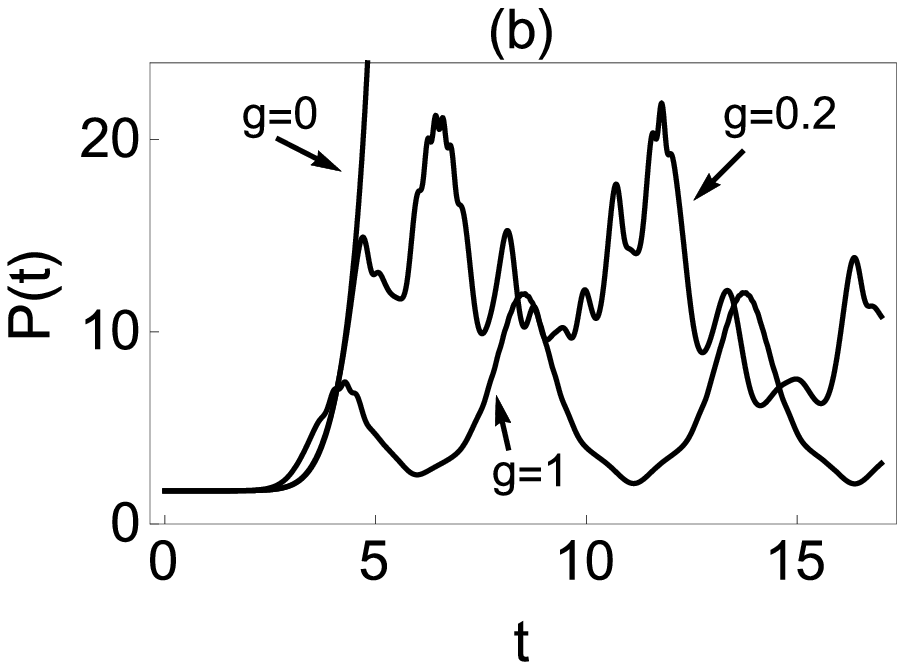}
\caption{ $U_1$  as a function of energy $E$ as obtained by numerically solving (\ref{2085sia2}) at $g=1$ (a) and $P(t)=\sqrt{u_1^2+u_2^2+u_3^2}$ ~ as a function of time as obtained by numerically solving (\ref{1ofhap2}) for some values of nonlinearity strength $g$ (b) for the nonlinear nonreciprocal trimer. The initial values for (b) are given by $u_1(0)=u_2(0)=1,u_3(0)=0.99$, where we introduce small amplitude modulation on $u_3$ (see the solution Equ. (10)). The amplitude grows unboundedly in the linear system $g=0$ as expected. Stability is restored for large values of $g$. }
\end{figure}

\section{Quadrimer}

Consider now that there are four sites in the system. Such a system is known as quadrimer. Taking into account conservative Kerr nonlinear term, the periodical system with asymmetric hopping amplitudes is governed by a system of the following dNLS equation
\begin{eqnarray}\label{oywsdfha?2}
i~\dot{u}_1= u_2-u_4+g~ |u_1|^2~ u_1 \nonumber\\
i~\dot{u}_2= u_3-u_1+g~ |u_2|^2 ~u_2 \nonumber\\
i~\dot{u}_3=u_4-u_2+g~ |u_3|^2 ~u_3 \nonumber\\
i~\dot{u}_4=u_1-u_3+g~ |u_4|^2 ~u_4 
\end{eqnarray}
where we set $\ds{\kappa=1}$ for the sake of simplicity. Despite the absence of the conservation of total intensity, this system can conserve the relative total population between the odd and even values of site numbers
\begin{equation}\label{vahdeklas}
\frac{d(|u_1|^2+|u_3|^2-|u_2|^2-|u_4|^2   )}{dt}=0
\end{equation}
Consider first the linear system $g=0$. In this case, all eigenvalues are purely imaginary since the system is anti-Hermitian and the corresponding eigenvectors form a Hilbert space. Therefore, any initial state either grows or decays in time as in the case of the linear nonreciprocal dimer model studied above. Consider next the nonlinear system $g\neq0$. In this case, stationary solutions are available. Let us look for the solutions of the form $\ds{u_{i}(t)=e^{-iE_i t} U_i}$ ( $E_3=E_1$ and $E_4=E_2$), where $\ds{E_1}$ and $\ds{E_2}$ are real-valued constants , $U_i$ are all time-independent constants. Therefore, we get
\begin{eqnarray}\label{oyrdhkja2}
E_1~U_1= U_2-U_4+g~ |U_1|^2~ U_1 \nonumber\\
E_2~U_2= U_3-U_1+g~ |U_2|^2 ~U_2 \nonumber\\
E_1~U_3=U_4-U_2+g~ |U_3|^2 ~U_3 \nonumber\\
E_2~U_4=U_1-U_3+g~ |U_4|^2 ~U_4 
\end{eqnarray}
We seek solutions of these equations by assuming that $U_i$ are real-valued constant. There are $4$ third-degree equations and hence there exists at most $3^4=81$ solutions. One of them is the trivial solution $U_1=U_2=U_3=U_4=0$ and $40$ of them are non-symmetric solutions since 
$  (-U_1,-U_2,-U_3,-U_4) $ are their symmetric solutions. At low values of the parameters $E_{1,2}$, there is a few branches of nonlinear modes (which coalesce at $E=0$) and bifurcation occurs as they are increased.  \\
The modes bifurcating from $0$ values are given by
\begin{equation}\label{dk29902j2}
U_1^2=U_3^2= \frac{E_1}{g}~,~~~U_2^2=U_4^2= \frac{E_2}{g}
\end{equation}
For small values of $E_1$ and $E_2$, there are no other solutions of (\ref{oyrdhkja2}). They are continuous modes depending on the  two parameters $E_1$ and $E_2$. Unfortunately, we numerically see that these solutions are all unstable against amplitude modulation as they grow unboundedly in time.\\
Below, we study our system when $\ds{E_1=E_2=E}$. We obtain solutions by classifying them according to whether some modes are dependent to other ones. Let us first start with the solutions where one of the mode amplitudes is fixed. There are two such solutions. The first one is given by $\ds{U_4=-U_2}$. In this case, our system is reduced to the nonlinear trimer model with open edges. Note that the set of the equation (\ref{2085sia2}) is written for the periodical boundary conditions, which does not lead to the conservation of the relative total population. The second such solution can be obtained if we assume $\ds{ U_3=-U_1}$. One can find at most $3^3=27$ solutions for these two reduced nonlinear trimers. The above solutions (\ref{dk29902j2}) with $E_1=E_2=E$ are also recovered in these solutions. The Fig. 4 plot $P=\sqrt{U_1^2+U_2^2+U_3^2+U_4^2}$ as a function of $E$ for these two solutions. There are a few modes at low $E$ and new branches appear as $E$ increases. For large values of $E$, there are $27$ modes, but most of them are very close to each other so it is difficult to distinguish them from the plots. We numerically study stability of these modes. We find that most of them grow exponentially in time if small amplitude modulations are introduced. \\
Having studied the case where only one of the modes is fixed, let us now look for the solutions where two of the modes are fixed. One can easily see that it is given by 
\begin{equation}\label{o67u?kja2}
U_3=-U_1~,~~~U_4=-U_2
\end{equation}
In this case, the quadrimer model is reduced to the dimer model with $\kappa=2$ (\ref{580sb2}). The two free parameters $U_1$ and $U_2$ are given by (\ref{5765utrfnc2}) and $\ds{U_2=2^{-1}(E-g~U_1^2) ~U_1}$, respectively. \\
Finally, we study the case where none of the mode amplitudes are equal to each other, $\ds{  U_1{\neq} U_2{\neq}U_3{\neq}U_4 } $. Unfortunately, we can't obtain exact analytical formulas for them. We numerically find that these modes are not stable and they grow unboundedly in time.  In Fig. 5, we plot $P=\sqrt{U_1^2+U_2^2+U_3^2+U_4^2}$ and the relative population $R=U_1^2+U_3^2-U_2^2-U_4^2$ as a function of $E$ for all solutions of (\ref{oyrdhkja2}) at $g=1$. 
\begin{figure}[t]
\includegraphics[width=6cm]{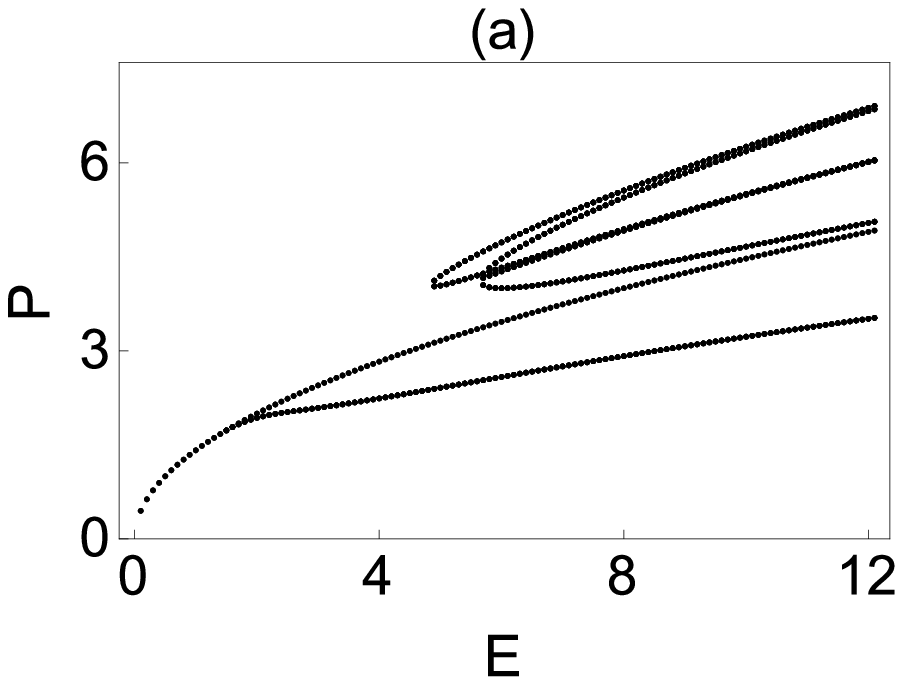}
\includegraphics[width=6cm]{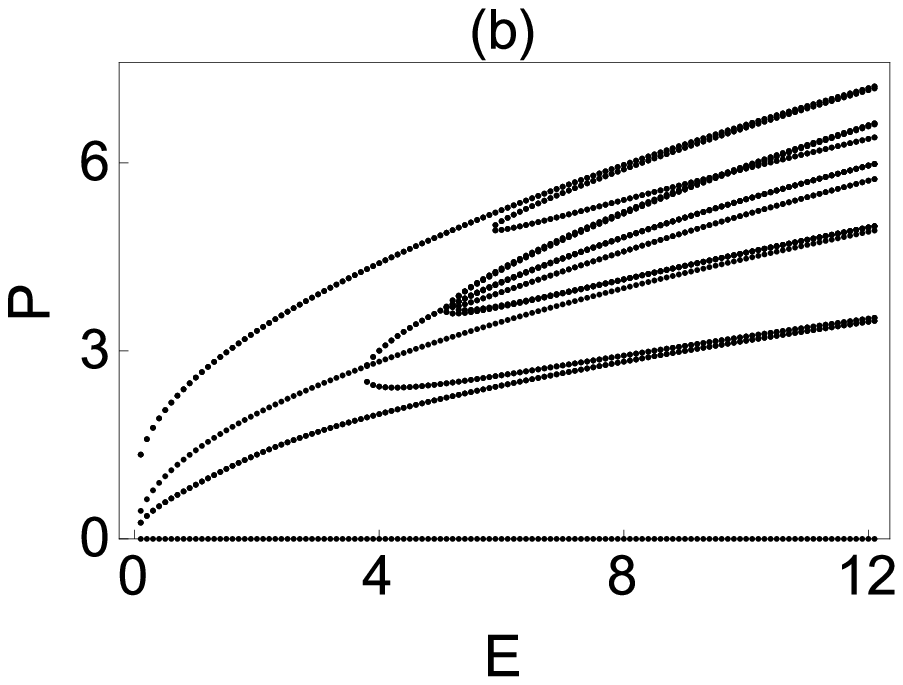}
\caption{ $P=\sqrt{U_1^2+U_2^2+U_3^2+U_4^2}$ as a function of energy $E$ as obtained by numerically solving (\ref{oyrdhkja2}) for $U_4=-U_2$ (a) and $U_3=-U_1$ (b) at $g=1$ for the nonlinear nonreciprocal quadrimer. There are at most 27 solutions, since the system is reduced to the trimer. Note that most of the modes have the same $P$. }
\end{figure}
\begin{figure}[t]
\includegraphics[width=6cm]{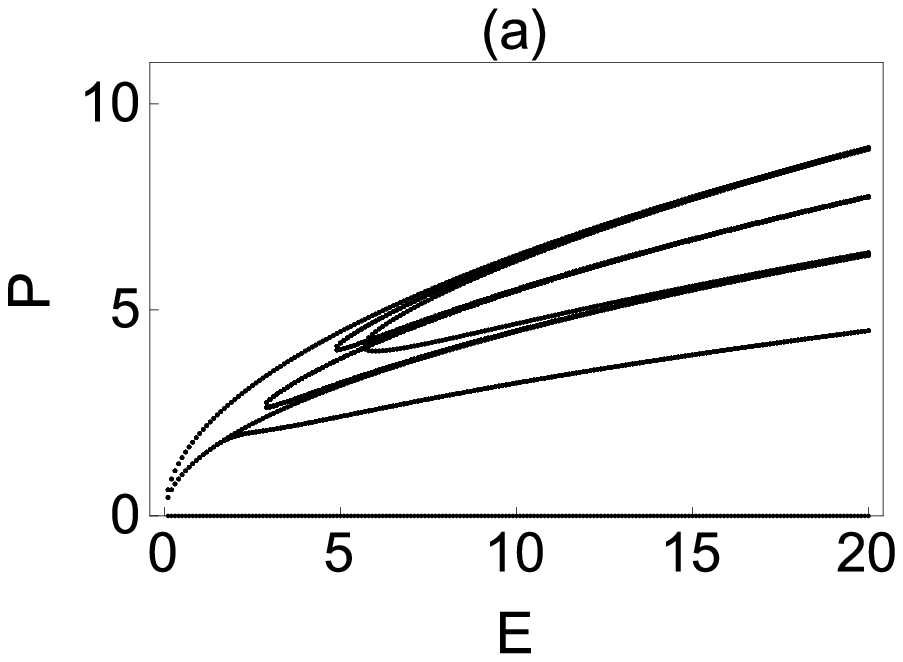}
\includegraphics[width=6cm]{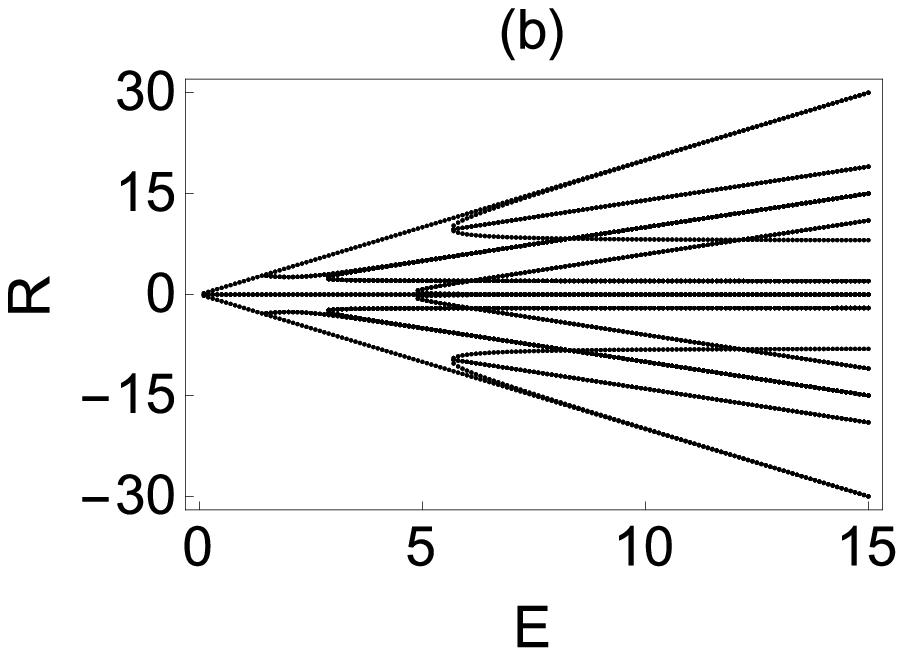}
\caption{ $P=\sqrt{U_1^2+U_2^2+U_3^2+U_4^2}$ and the relative population $R= U_1^2+U_3^2-U_2^2-U_4^2$ as a function of energy $E$ at $g=1$. We assume that $\ds{  U_1{\neq} U_2{\neq}U_3{\neq}U_4 } $ for the numerical solutions of the Equ. (\ref{oyrdhkja2}). These modes are not stable.}
\end{figure}\\
We have seen that the nonlinear quadrimer problem can be reduced to the nonlinear dimer one. A question arises. Can we make such a reduction for a non-reciprocal nonlinear lattice with large number of lattice sites? Consider an array of $4N$ waveguides coupled to each other asymmetrically, where N is a positive integer. For $N=1$, the system corresponds to the quadrimer as described above.  Assume that the system has periodical boundary condition. The system is governed by
\begin{eqnarray}\label{oodas?l}
i~\dot{u}_1&=& u_2-u_{4N}+g~ |u_1|^2~ u_1 \nonumber\\
i~\dot{u}_n&=& u_{n+1}-u_{n-1}+g~ |u_n|^2~ u_n \nonumber\\
i~\dot{u}_{4N}&=&u_1-u_{4N-1}+g~ |u_{4N}|^2 ~u_{4N} 
\end{eqnarray}
where $n=2,3,..,4N-1$. One can show that the relative total population between the odd and even values of site numbers is conserved in time
\begin{equation}\label{vahdeklas}
\frac{d( \sum_{n=1}^{4N}  ( |u_{2n-1}|^2-|u_{2n}|^2  ) )}{dt}=0
\end{equation}
We can reduce this system as a noninteracting collection of the nonlinear dimer by assuming
\begin{eqnarray}\label{78995s}
u_{n+2}=(-1)^n ~u_{n}
\end{eqnarray}
where $n=1,2,..,4N-2$. The corresponding stationary nonlinear modes can be found in the section 2.

\section{Conclusion}

It is well known that stationary modes appear in nonlinear nonconservative systems. In the literature, nonlinear systems with gain and loss have been mostly explored and nonreciprocal nonlinear systems have not been understood yet. Here, we have considered nonlinear extension of anti-Hermitian systems. A linear anti-Hermitian Hamiltonian has purely imaginary eigenvalues and orthogonal eigenvectors, which implies that Hilbert space can be formed. Therefore no stationary state is available in linear anti-Hermitian systems. We have constructed stationary nonlinear modes and studied their stability properties. We have shown that the nonlinear nonreciprocal dimer do not conserve total intensity but relative population as opposed to parity-time symmetric non-Hermitian lattices. Therefore, no particle transfer occurs between the sites and the system either grows unboundedly or oscillate as a whole. We have extended our analyses to nonreciprocal trimer and quadrimer.

\end{document}